\documentclass[11pt,twoside]{article}
\usepackage{cspm2015}

\aspSuppressVolSlug
\resetcounters

\def\figspath{.} 

\begin{document}

\title{Ground-based Observations of the Solar Sources of Space Weather (Invited Review)}
\author{Astrid~M.~Veronig$^{1,2}$ and Werner~P\"otzi$^2$}
\affil{$^1$Institute of Physics/IGAM, University of Graz, Austria}
\affil{$^2$Kanzelh\"ohe Observatory for Solar and Environmental Research, University of Graz, Austria}

\paperauthor{Astrid M. Veronig}{astrid.veronig@uni-graz.at}{ORCID_Or_Blank}
  {Institute of Physics/IGAM}{University of Graz}{Graz}{}{8020}{Austria}
\paperauthor{Werner P\"otzi}{werner.poetzi@uni-graz.at}{0000-0003-4811-0543}
  {Kanzelh\"ohe Observatory for Solar and Environmental Research}{University of Graz}
  {Treffen}{}{9521}{Austria}

\begin{abstract}
Monitoring of the Sun and its activity is a task of growing importance in the 
frame of space weather research and awareness. Major space weather disturbances 
at Earth have their origin in energetic outbursts from the Sun: solar flares, 
coronal mass ejections and associated solar energetic particles.
In this review we discuss the importance and complementarity of ground-based and 
space-based observations for space weather studies. The main focus is drawn on 
ground-based observations in the visible range of the spectrum, in particular 
in the diagnostically manifold H$\alpha$ spectral line, which enables 
us to detect and study solar flares, filaments (prominences), filament 
(prominence) eruptions, and Moreton waves. Existing H$\alpha$ networks such as 
the GONG and the Global High-Resolution H$\alpha$ Network are discussed. As an 
example of solar observations from space weather research to operations, we present 
the system of real-time detection of H$\alpha$ flares and filaments established 
at Kanzelh\"ohe Observatory (KSO; Austria) in the frame of the space weather segment 
of the ESA Space Situational Awareness programme (\url{swe.ssa.esa.int}). 
An evaluation of the system, which is continuously running since July 2013 is 
provided, covering an evaluation period of almost 2.5 years. During this period, 
KSO provided 3020 hours of real-time H$\alpha$ observations at the ESA SWE portal. 
In total, 824 H$\alpha$ flares were detected and classified by the real-time 
detection system, including 174 events of H$\alpha$ importance class 1 and larger. 
For the total sample of events, 95\,\% of the automatically determined flare peak 
times lie within $\pm$5 min of the values given in the official optical flares 
reports (by NOAA and KSO), and 76\,\% of the  start times. The heliographic positions 
determined are better than $\pm$5$^\circ$. The probability of detection of flares 
of importance 1 or larger is 95\,\%, with a false alarm rate of 16\,\%. These 
numbers confirm the high potential of automatic flare detection and alerting from 
ground-based observatories.
\end{abstract}

\section{Introduction}

Space weather refers to variations in the space environment of our solar system 
driven by the Sun that can affect technologies in space and  on Earth. The main 
sources of severe space weather disturbances are solar flares,  coronal mass 
ejections (CMEs) and solar energetic particles (SEPs). CMEs are large-scale 
ejections of magnetized plasma from the Sun to interplanetary (IP) space with speeds 
of some hundred up to 3000~km~s$^{-1}$ \cite[{\it e.g.} reviews by][]{forbes2006,chen2011,webb2012}. 
Earth-directed CMEs with a strong negative $B_z$ component, {\it i.e.}  oppositely directed to the Earth's magnetic field, 
cause magnetic reconnection at the day-side magnetosphere. In this process, energy 
and particles from the solar plasma couple to the Earth magnetosphere, thus causing 
a geomagnetic storm and induced geo-electric fields. Flares are sudden enhancements 
of the solar radiation, most significantly at long (radio) and  short (EUV, X-ray) 
wavelengths \cite[{\it e.g.}, reviews by][]{Benz2008,Fletcher2011}.  The enhanced 
EUV and X-ray radiation provides additional heating of the Earth's outer atmosphere. As a result, the 
atmosphere expands, thus increasing the local density and the aerodynamic drag on satellites in low-Earth orbits. 
CMEs and flares are both capable of accelerating Solar Energetic Particles (SEPs). 
SEPs pose a severe hazard to astronauts in space due to their ionising radiation, 
and they may also disrupt technological systems and electronics on-board satellites 
(see, {\it e.g.}, \cite{bothmer2007}, \cite{Schwenn2006} and \cite{Pulkkinen2007}).

A variety of measures and actions are currently undertaken to assess the risks 
and to ultimately prevent us from severe effects of strong space weather events, 
see {\it e.g.} ESA's {\it Space Situational Awareness} (SSA) Programme and the U.S.\ 
{\it National Space Weather Strategy} and {\it National Space Weather Action Plan} 
documents  released in October 2015\footnote{\url{www.swpc.noaa.gov/news/national-space-weather-strategy-and-action-plan-released}}. 
Since the ultimate origin of space weather 
disturbances is our Sun, it is important to better understand the physics of these 
processes and to continuously monitor the Sun and its activity. In this review, 
we discuss the observations of the Sun relevant to space weather. The focus lies 
on the observations of solar flares and eruptions from ground-based observatories, 
with special emphasis on the diagnostics potential of the H$\alpha$ spectral line. 
As an example, we present the real-time detection system of solar flares and 
filaments at Kanzelh\"ohe Observatory, which was recently established in the 
frame of the space weather segment of ESA's SSA programme.

\section{Solar observations for space weather: ground-based versus space-based}

\subsection{Solar key observations for space weather research and monitoring}

Key observations of the Sun with respect to space weather include the solar magnetic 
field, which is the ultimate source of all space weather events, together with 
data capturing the dynamics of the solar corona, chromosphere and photosphere. 
The vast amount of energy that is impulsively released in flares and CMEs is originally stored 
in the kilo-gauss fields of sunspots. Turbulent plasma motions in the convection 
zone and photosphere shear and twist the coronal magnetic field, thus building 
up electric currents in the corona \cite[{\it e.g.}\ reviews by][]{priest2002,wiegelmann2014}. 
The {\it free} energy stored in these currents may eventually be released, triggered 
by some instability. Whereas the energy build-up occurs gradually over days to 
weeks, the energy release via flares and CMEs occurs impulsively, {\it i.e.}\ on the 
order of minutes to hours. Full-disk magnetic field measurements of the Sun are 
performed by several ground-based observatories. However, the most regular 
observations are currently provided by the Helioseismic Magnetic Imager (HMI) 
onboard the Solar Dynamics Observatory (SDO), with a cadence of 1~min for 
the line-of-sight (LOS) component and 12~min for the full-vector field at a spatial 
resolution of about 1~arcsec. 

Due to its high temperature, the solar corona emits predominantly at soft X-ray (SXR)
and Extreme Ultraviolet (EUV) wavelengths. Since the emission at these wavelengths 
is absorbed by the Earth's atmosphere, the coronal dynamics and evolution of solar 
flares can only be observed by space-based EUV or SXR telescopes. Currently, regular 
high-cadence observations of the full Sun in the EUV are performed by the Atmospheric 
Imaging Assembly (AIA) onboard SDO and the SWAP instrument onboard Proba2. 
Observations of the propagation of CMEs up to some ten solar radii through the corona, 
which are necessary in order to estimate their initial speed and direction of 
motion close to the Sun, is also restricted to observations from space. 
Coronagraphs block the direct emission from the solar photosphere in order to 
measure the faint emission that is Thomson-scattered at free electrons in the highly 
ionized corona. Currently, the Large Angle and Spectrometric Coronagraph (LASCO) 
on-board the Solar and Heliospheric Observatory (SOHO) provides regular observations 
of CMEs from the Sun-Earth line. In addition, STEREO-A (and the recently lost 
STEREO-B) COR1 and COR2 coronagraphs observe CMEs from varying positions away 
from the Sun-Earth line, providing us with an alternative vantage point to observe
CMEs and to reconstruct their 3D evolution. The propagation of CMEs in IP space 
up to the distance of Earth and beyond is tracked by the wide-field cameras 
of the Heliospheric Imagers (HI) onboard the STEREO spacecraft. 

In contrast to the solar corona, the solar chromosphere can be well observed by 
Earth-based observatories, especially in strong Fraunhofer absorption lines in 
the visible wavelength domain.\footnote{ The other important spectral window 
accessible from ground in addition to the visible part of the spectrum is the 
radio domain. Radio heliographs and radio spectrographs provide an important means 
for diagnosing, {\it e.g.}, flare-accelerated high-energy electrons escaping along open field lines to IP space 
(type III burst) and outward moving shock waves produced by fast CMEs (type II bursts). In this review 
we concentrate on ground-based observations in the visible light. For reviews on 
solar radio physics and its relation to space weather we refer the interested 
reader to \cite{Gary2004,Pick2006,Pick2008}.}    
Examples include the H$\alpha$ line (656.28~nm) of the neutral hydrogen atom and the 
Ca\,{\sc ii}\,H (396.85~nm) and K (393.37~nm) lines of ionised calcium. The 
H$\alpha$ line has been the main spectral window for regular observations 
of solar flares since the 1950s, in particular during and in the follow-up of the 
International Geophysical Year (IGY) in 1957. The first flare classification scheme 
introduced and still in use in parallel to the GOES soft X-ray classification is 
the H$\alpha$ importance scheme. It is based on the area and the brightness of solar flares in 
the H$\alpha$ spectral line \citep{Zirin1988}. The importance classes ranging 
from the smallest to the largest events are subflares, flares of importance 1, 2, 3, 
and 4. This importance class is complemented by a character signifying the 
H$\alpha$ line center brightness enhancement of the event as f(aint), n(ormal) 
or b(rilliant).  

In addition to the observations of the chromospheric dynamics of solar flares, 
H$\alpha$ data provide also insight into the activation and eruption of solar 
prominences (filaments), which are often associated with coronal mass ejection 
\cite[{\it e.g.},][]{gopal2003,bein2012}. Prominences are thought to outline the 
inner part of a magnetic flux rope, where the cool prominence material, suspended 
in the corona, is balanced by the upward directed Lorentz force in the dips of 
the helical flux rope. During the CME lift-off, the prominence may become visible 
as the core of the three-part CME structure \citep{Illing1985} in 
the coronagraph images. In addition, H$\alpha$ image sequences may also reveal the 
motion of Moreton waves. Moreton waves are the chromospheric ground-track of fast-mode 
MHD shock waves that propagate through the solar corona, most likely initiated 
by the impulsive initial expansion of a CME. The pressure gradient at the coronal shock front
compresses and pushes the chromospheric plasma downward, which becomes visible 
as a bright propagating arc in the center of the H$\alpha$ line  
\cite[see, {\it e.g.}, reviews by][]{vrsnak2008,warmuth2015}. Due to the LOS motion of the 
chromospheric plasma of the order of 10~km~s$^{-1}$, Moreton waves are best visible 
in the wings of the H$\alpha$ spectral line.

\subsection{Ground-based versus space-based}

Solar observations for space weather research and monitoring have to cover the 
full solar disk and to be performed with high cadence (say $\lesssim$1~min) in order to 
capture the dynamics of the events. The requirements on spatial resolution are 
moderate, in the order of $\lesssim$3$''$. In addition, the observations have to be 
available in almost real-time. 

The main cons of ground-based observations against space-based observations are the following:
\begin{itemize}
\item atmospheric issues: weather and seeing negatively influence the availability 
  of the observations as well as the quality of the images;
\item day-night cycle, {\it i.e.}\ at each individual ground-based station the 
  observation time is limited; 
\item few wavelength windows are accessible from ground: visible, radio, near IR.
\end{itemize}
However, there are also a number of pros for ground-based observations compared 
to space-based observations:
\begin{itemize}
\item relatively ``simple" and inexpensive observing systems are needed;
\item flexibility in changing and upgrading the system (filters, hardware, etc.);
\item several strong Fraunhofer lines are available in the visible range with 
  good diagnostics potential for flares and solar eruptions; 
\item no telemetry constraints, thus high-cadence observations and analysis can be easily performed; 
\item no telemetry delays, thus the data are accessible in real-time;
\item networks of observatories can overcome the limitations in observing time of individual sites.
\end{itemize}

\subsection{Networks for ground-based solar monitoring}

The first network for the observation of the Sun with respect to its activity and 
space weather effects was actually established in the early 1940s during WW\,II 
by Karl-Otto Kiepenheuer under the authority of the ``Deutsche Luftwaffe'' (German Airforce). 
In the 1930s it was discovered that the state of the Earth's ionosphere 
and thus the propagation characteristics of radio waves is affected by the activity 
of the Sun \citep{Dellinger1935}. During WW\,II, short-wave radio transmission was an 
important means of communication and navigation, in particular for the airforce 
and the marine. Thus, the German Airforce founded a network of observatories in 
the Alps in order to regularly monitor the Sun and to quantify its activity. 
The goal was to inform the German Airforce in case of ionospheric disturbances and 
to provide forecasts of such disturbances. The network consisted of three stations in 
Germany (Zugspitze, Schauinsland and Wendelstein) and Kanzelh\"ohe Observatory in 
Austria \citep{Seiler2007,Jungmeier2014a}. 

A great leap forward for the observations and studies of solar-terrestrial relations 
was the International Geophysical Year (IGY) 1957--1958. In the context of the IGY, 
a world-wide network of ground-based observatories was established to observe solar 
H$\alpha$ flares.  Since the end of the year 1955, data from a variety of geophysical 
and solar observatories worldwide were compiled in the Solar-Geophysical Data (SGD) 
bulletins (until the year 2009). Current ground-based networks of observatories to 
monitor solar activity are the {\it Global High-Resolution H$\alpha$ Network} led 
by the New Jersey Institute of Technology 
\citep[formerly by Big Bear Observatory;][]{Steinegger2000} and the H$\alpha$ 
observations by the NSO {\it Global Oscillation Network Group} 
(GONG; \citeauthor{Hill1994} \citeyear{Hill1994}).  

The GONG network was established in order to study the solar interior by the means 
of helioseismology. The network consists of six observing sites around the world, 
well distributed in longitude in order to basically cover 24 hours of observations 
per day \citep{Hill1994}. The sites are:  High Altitude Observatory at Mauna Loa 
(Hawaii, USA; W156$^\circ$), Big Bear Solar Observatory (California, USA; W117$^\circ$), 
Cerro Tololo Interamerican Observatory (Chile; W71$^\circ$), Observatorio del Teide 
(Canary Islands, Spain; W17$^\circ$), Udaipur Solar Observatory (India; E74$^\circ$) 
and Learmonth Solar Observatory (Australia; E114$^\circ$). 
The GONG observations began in the year 1994 and the network is able to provide solar observations
for 96\,\% of the time. In the year 2010, the GONG network stations were 
complemented by H$\alpha$ telescopes \citep{Harvey2011}. The H$\alpha$ system 
is an add-on to the normal GONG helioseismology instrument. 
A beamsplitter sends 656~nm light to a mica etalon filter with a 
full-width at half-maximum (FWHM) of 0.04\,nm. The H$\alpha$ images are  
recorded by a CCD with 2048$\times$2048 pixels. Automatic exposure control is 
in place that maintains the quiet center of the solar disk at 20\,\% of the full 
dynamic range available in order to avoid saturation during flares. H$\alpha$ images 
are collected once a minute, immediately corrected for flat field and dark current, 
and sent to a central archive where they are available in quasi real-time (\url{halpha.nso.edu}).

The Global High-Resolution H$\alpha$ Network (GHN) utilizes already existing 
facilities at nine observatories around the world, that have each good seeing 
conditions, a high coverage of sunny days over the year and adequate observing staff. 
In contrast to the GONG network, the H$\alpha$ instruments of the GHN stations 
are not identical. The nine stations of the GHN are: Big Bear Solar Observatory 
(California, USA; W117$^\circ$), Catania Astrophysical Observatory (Italy; E15$^\circ$), 
Kanzelh\"ohe Observatory (Austria; E14$^\circ$), Uccle Solar Equatorial Table 
(Belgium; E4$^\circ$), Observatory de Paris-Meudon (France; E2$^\circ$),
Observatoire Midi-Pyr\'en\'ees (France; E0$^\circ$), Yunnan Solar Observatory 
(China; E103$^\circ$), and Huairou Solar Observing Station (China; E117$^\circ$). 
The sizes of the GHN H$\alpha$ images are between 1024$\times$1024 and 
2048$\times$2048 pixels, and the observing cadence at each site is at least one 
image per minute. The goal of the GHN network is to cover the night gap in order
to provide data for filament eruption detection, solar rotation measurements and 
flare prediction algorithms \citep{Steinegger2000,Steinegger2001}. The data of 
the individual stations is collected at the Space Weather Research Lab of the 
New Jersey Institute of Technology (\url{swrl.njit.edu/ghn_web}).

\articlefigure[width=0.9\textwidth]{\figspath/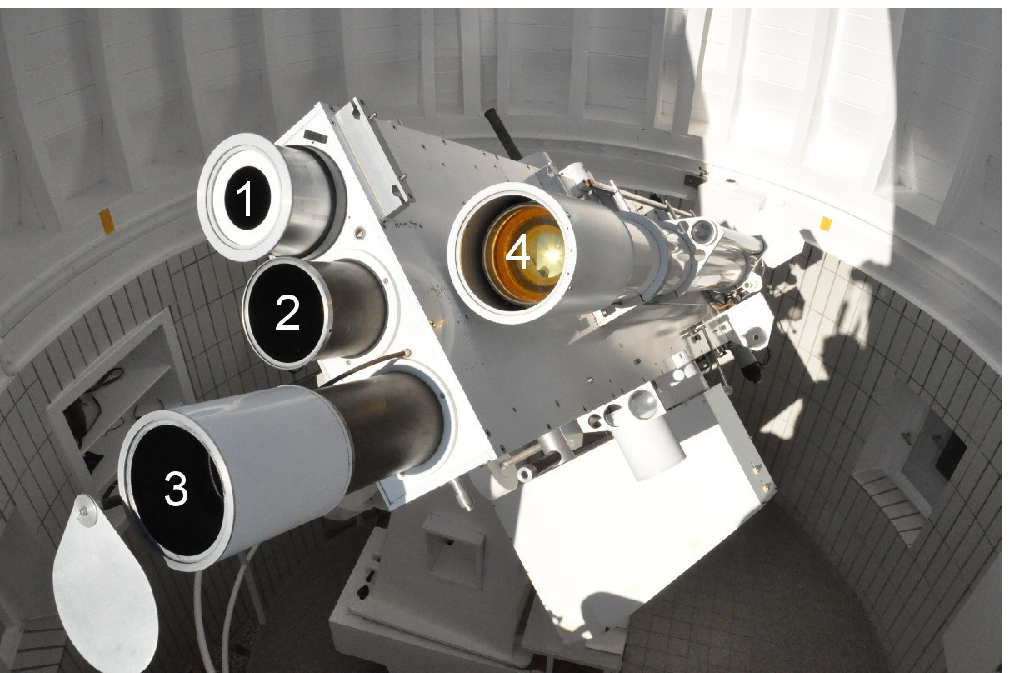}{telescope_fig}
{The KSO surveillance telescope consisting of three refractors observing the Sun in 
Ca\,{\sc ii}\,K (1), H$\alpha$ (2) and  white-light (4) plus an additional refractor 
producing a projection of the solar disc for sunspot drawings (3).}

\section{Real-time flare and filament detection at Kanzelh\"ohe Observatory}

In this section we describe the automatic system of real-time detection of solar 
flares and filaments that was developed and implemented in 2012/2013 at 
Kanzelh\"ohe Observatory. This activity is part of the space weather segment of 
ESA's SSA programme, and aims at monitoring and automatic detection of the solar 
sources of space weather events (flares, filament eruptions) in quasi real-time 
using ground-based H$\alpha$ imagery. In addition, we provide a preliminary 
assessment of the results of this system since its establishment, covering a 
period of 29 months from 7/2013 to 11/2015. A detailed description of the overall 
system, its implementation and an evaluation of its first 5~months running 
in real-time is given in \cite{Poetzi2015}.

\subsection{KSO instrumentation and observations}

Kanzelh\"ohe Observatory for Solar and Environmental Research 
(KSO; \url{kso.ac.at}; N 46$^\circ$40.7$'$, 13$^\circ$54.1$'$, 
altitude 1526 m) of the University of Graz (Austria) regularly performs
high-cadence full-disk observations of the Sun in the 
H$\alpha$ spectral line \citep{Otruba2003}, the Ca\,{\sc ii}\,K spectral line 
\citep{Polanec2011}, and in white-light \citep{Otruba2008} with a coverage of 
about 300 observing days a year. All telescopes are mounted onto the 
KSO surveillance telescope (Figure \ref{telescope_fig}), and the data are publicly 
available {\it via} the online KSO data archive at \url{cesar.kso.ac.at} 
\citep{Poetzi2013}.

\articlefigure[width=0.87\textwidth]{\figspath/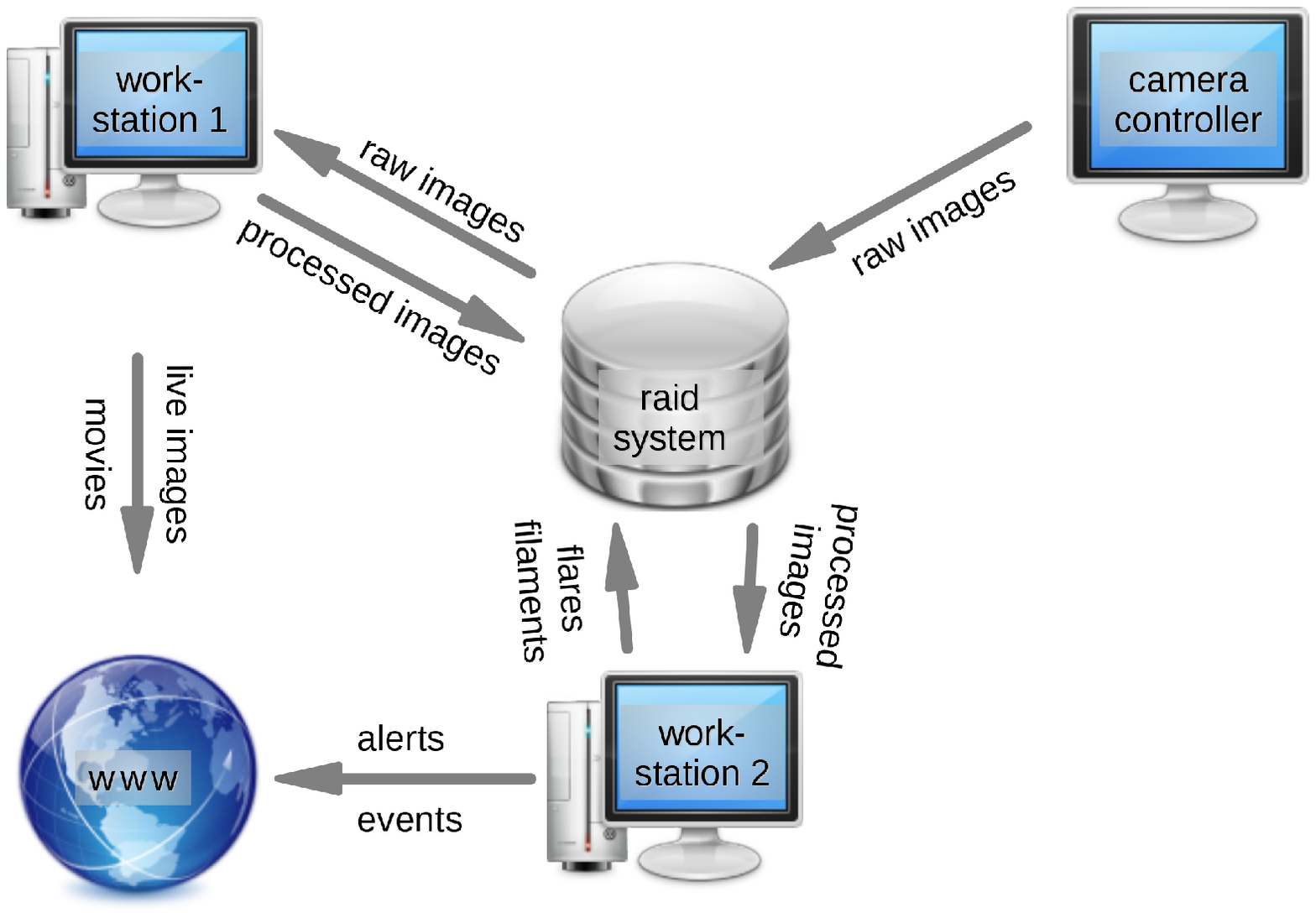}{proc_fig}
{Three computers are involved in the processing pipeline of the KSO H$\alpha$ 
observing system: the camera controller and two workstations. 
All computers work in parallel. The data and the results from the image 
recognition algorithm are channeled through and saved at a central raid system.}

The KSO H$\alpha$ telescope is a refractor with an aperture ratio number 
of $d/f = 100/2000$ and a Lyot band-pass filter centered at the H$\alpha$ spectral 
line with a FWHM of 0.07\,nm. The H$\alpha$ images are 
recorded by a CCD of 2048\,$\times$\,2048 pixels (12-bit dynamic range) with a spatial resolution of about 1$''$
and a cadence of 6~s. A frame rate 
of seven images per second permits the application of frame selection 
in order to profit from moments of good seeing conditions. An automatic 
exposure control system is in use in order to avoid saturation effects in strong flares.

\subsection{The KSO H$\alpha$ real-time flare and filament detection system}

To optimize the process and speed of the real-time data provision as well as 
the automatic flare and filament detection, different computers are connected to 
the KSO H$\alpha$ observing system that run in parallel, each one performing a 
specific set of tasks: i) the camera computer is responsible for image acquisition, 
ii) workstation 1 performs the quality check of the incoming images, the data 
reduction and online data provision, iii) workstation 2 performs the image 
recognition including the automatic flare detection, filament detection 
and event alerting (see Figure \ref{proc_fig}).

Each H$\alpha$ image is grabbed by the camera computer and sent to workstation 1, 
where the image is checked for its quality. The quality test is based on three 
main parameters, the accuracy of solar radius detection, the large-scale intensity 
distribution and the image sharpness. If an image passes the quality criterion, 
it is processed and published on the web server. In parallel, the processed 
image is also transferred to workstation 2, on which the image recognition 
algorithm is performed. If an event is detected, its characteristic parameters 
are calculated. In case that the event exceeds a certain threshold 
({\it i.e.}\  flare area/importance class), a flare alert is published online 
at ESA's SSA SWE portal and an alert e-mail is sent out.

\articlefigure[width=0.87\textwidth]{\figspath/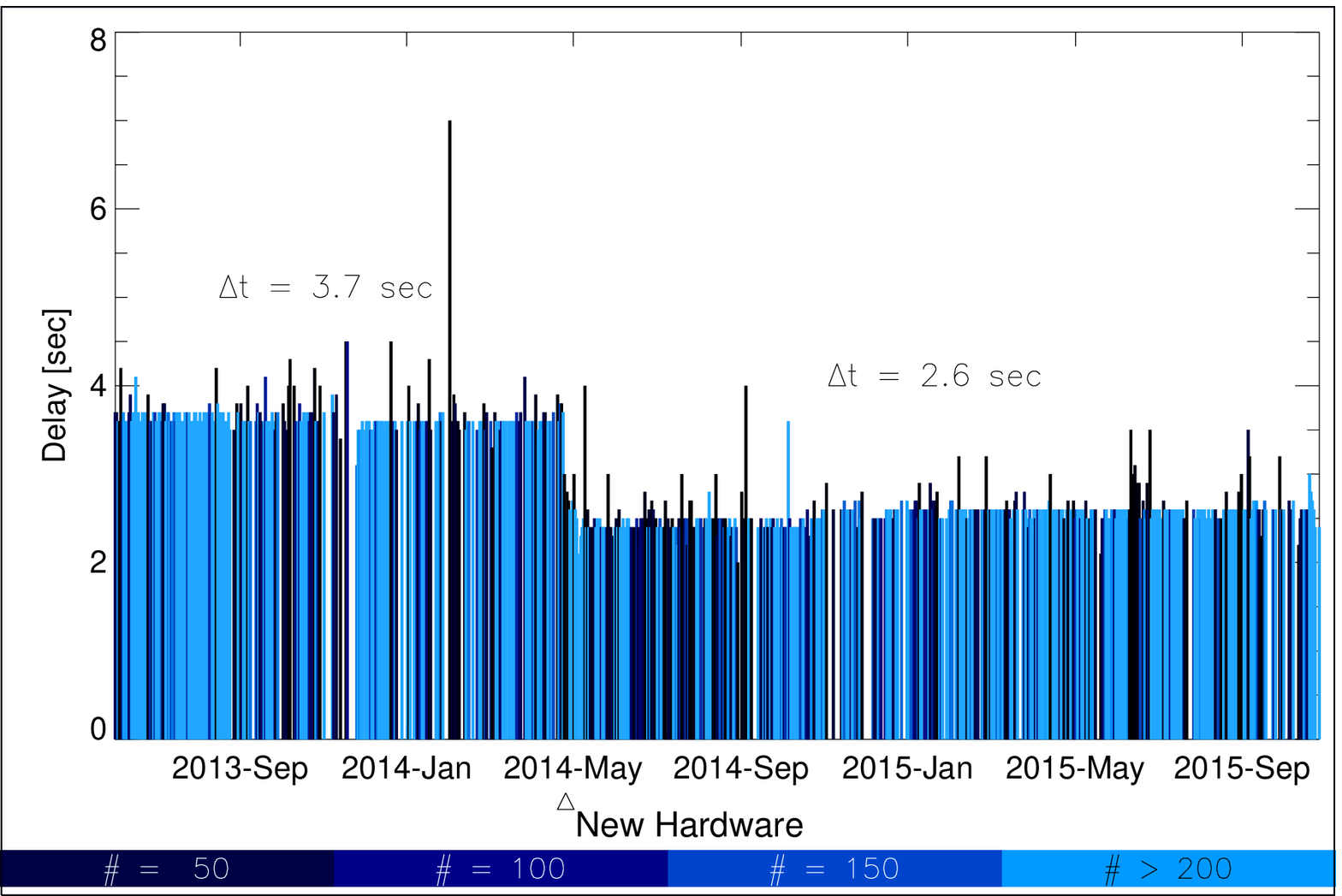}{delay_fig}
{Time delay between the capture of an H$\alpha$ image and its online provision on the 
ESA SSA SWE portal. Dark bars denote days with only a few observations and 
bright bars denote good observation days ({\it cf.}\ color bar at the bottom). 
An upgrade in the hardware components in May 2014 led to a further improvement 
of the performance of the system.}

\subsubsection{Provision of real-time images and movies on the ESA SSA SWE portal}
 
Each minute an image is selected for the real-time H$\alpha$ display at 
ESA's SSA SWE portal (\url{swe.ssa.esa.int/web/guest/kso-federated}). 
The size of the image is reduced to 1024$\times$1024 pixels and stored in jpeg 
format for fast and easy display. The image is overlaid with a solar coordinate 
grid and annotated with a header containing the time information. In addition, 
the H$\alpha$ images recorded during the latest hour of observations  are animated 
via an html movie at the SWE portal. For later validation, a log file that keeps 
track of the image acquisition time and the time when the image was provided 
online is updated for each image. 

The online data provision for the ESA SSA SWE portal worked almost without any
interruptions, only 55 out of a total of 181.221 images during the period July 2013 to November 
2015 did not appear on the portal, corresponding to 0.03\%. In Figure~\ref{delay_fig} 
we show for each day the mean time delay between the capture of an image and its 
online provision at the portal. At the beginning this delay was about 3.8~s, and 
then improved even further to 2.6~s thanks to a hardware upgrade.

\subsubsection{Real-time flare detection and alerting}
 
Many of the iterative algorithms of the image recognition are computationally 
intensive. However, they can be easily parallelized and thus one can efficiently
utilize the computational power of modern graphic processing units (GPU).
The image recognition algorithm has been implemented in C\raisebox{0.5ex}{\tiny\textbf{++}} 
and installed on a dedicated machine with a high-performance GPU. The system  benefits from the large number of processing 
units which are used for the highly parallelized computations. The algorithm needs less than 10~s to process the flare and filament recognition 
on one $2048 \times 2048$ pixel image, which allows event detection in near real-time. 
The results of the image recognition algorithm are stored in log files 
containing tables of the flares and filaments detected. The log files  
are updated with each new image that enters the processing pipeline, and are then used us to derive 
the evolution of the detected features. Detailed descriptions of the image algorithms developed and their implementation 
at the KSO H$\alpha$ observing system are given in \cite{Riegler2013a,Riegler2013b,Poetzi2015}.

For the detection of flares and filaments in the KSO real-time data, a 
segmentation of every H$\alpha$ image into four classes is conducted, namely the classes 
{\it filament}, {\it flare}, {\it sunspot} and {\it background}. The main characteristics 
used in the image segmentation is the intensity and the shape of the objects. 
The image recognition algorithm developed consists of four main building blocks. 
(i) The preprocessing handles the different intensity distributions, large-scale 
inhomogeneities and noise with the two goals of image normalization and feature enhancement. 
(ii) The feature selection step defines the characteristic attributes of the features of interest, 
i.e.\ flares and filaments, and how to model them. The data used for feature
extraction and learning of the model were derived from a set of KSO H$\alpha$ images, 
where an expert labeled the different classes. (iii) In the multi-label 
segmentation step, the model derived is applied to the real-time images. 
(iv) In the postprocessing, every object that was identified gets a unique ID assigned. 
This allows us to track the detected features in the H$\alpha$ image 
sequence, and to compute the evolution of flares, filaments and filament eruptions. The 
segmentation results are then used to derive characteristic properties of the 
objects identified, which are needed for the subsequent classification of the events.

The characteristic properties for flares include the heliographic 
position, the flare area (which defines the importance class), the brightness 
class, and the flare start, peak and end times. These quantities need not only the 
information of a single H$\alpha$ image but also the information stored in the 
log files for the previous time steps. Handling of simultaneous 
flares is easily possible as each flare is identified {\it via} a unique ID that is 
propagated from image to image. In Figure~\ref{flare_fig} we show as an example 
the evolution of the area and the brightness for a 2B class flare that occurred 
on August 22, 2015 (top panel). The bottom panels show the original H$\alpha$ image 
during the peak of the event (left) and the detected flare area (right).

\articlefigure[width=0.89\textwidth]{\figspath/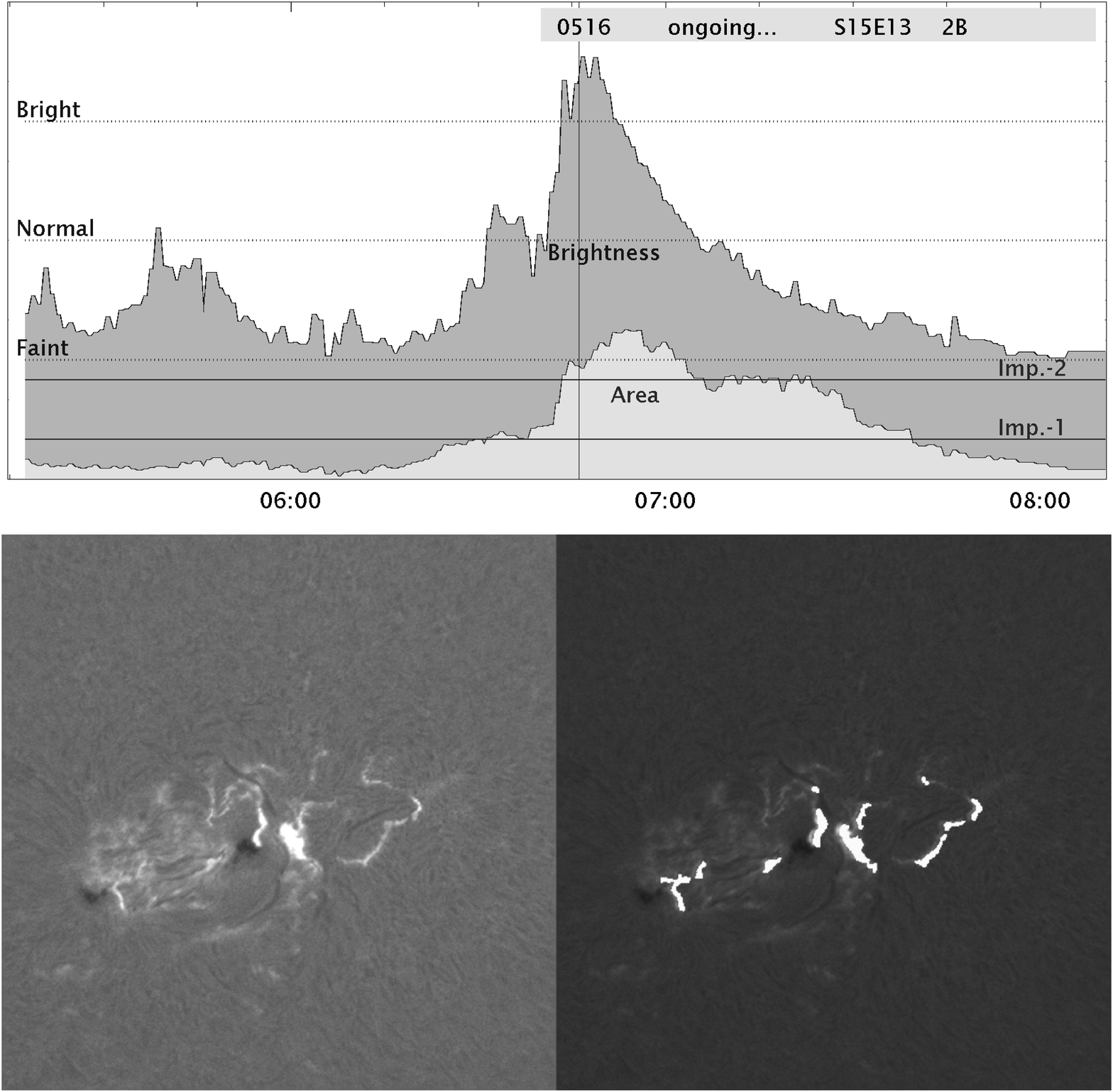}{flare_fig}
   {Illustration of the flare parameter calculation for a sample 2B flare that occurred on 
   August 22, 2015.  Top: Evolution of the flare area (light grey) and flare brightness 
   (dark grey) determined in real-time by the KSO automatic flare detection system. 
   Bottom: snapshot of the flare at peak time (indicated by the vertical line in the 
   top panel) together with the instantaneous flare detections 
   (white areas; right panel). The inset in the upper right corner shows the 
   start time, the peak time or status (``ongoing''), the heliographic position and 
   the flare importance class derived. The information shown in the inset are 
   also made available in real-time on the ESA SSA SWE homepage, 
   \url{swe.ssa.esa.int/web/guest/kso-federated}.
   }

The flare area is calculated by the number of segmented pixels with the same ID. 
These are subsequently converted by the pixel-to-arcsec scale of that day to derive the 
area in millionths of the solar hemisphere. The conversion procedure includes the 
information of the heliographic flare position in order to correct for foreshortening 
toward the solar limb. The determined area is then directly converted to the flare 
importance class (subflares, 1, 2, 3, 4) according to the official flare importance 
definitions \citep{Zirin1988}. For the categorization into the flare brightness 
classes (Brilliant-Normal-Faint: B-N-F), the intensity values relative to the 
background are calculated.

For the automatic flare detection and characterization, several definitions are applied:
The \emph{flare start} is defined as the time when the brightness enhancement was detected above 
the faint flare level for three consecutive images. The \emph{peak time} of the flare 
is defined as the time where the maximum flare brightness is reached. The \emph{flare 
position} is defined by the location of the brightest flare pixel in the normalized 
and filtered H$\alpha$ images at the time of the flare peak. The \emph{importance class} 
of the flare is defined {\it via} the maximum area of the flare, and is updated 
when the area exceeds the level of a higher importance class. The \emph{flare end} 
is defined as the time when the brightness has decreased below the faint level 
for ten consecutive images or when there is a data gap of more than 20 minutes. 

If a flare that exceeds a certain threshold is detected, {\it i.e.}\  a certain 
importance class, then a flare alert is published on the ESA SSA SWE portal and 
an alert email is sent out to registered users. At the same time the event list 
on the ESA SSA SWE portal is updated. Further updates are made when more 
information on a flare becomes available during its evolution ({\it e.g.}, the peak time) 
or when a flare, that is already listed, increases in its importance class. 
A list of all detected events is stored in the local raid system for later 
evaluation and preparation of an archive.

\articlefigure[width=0.87\textwidth]{\figspath/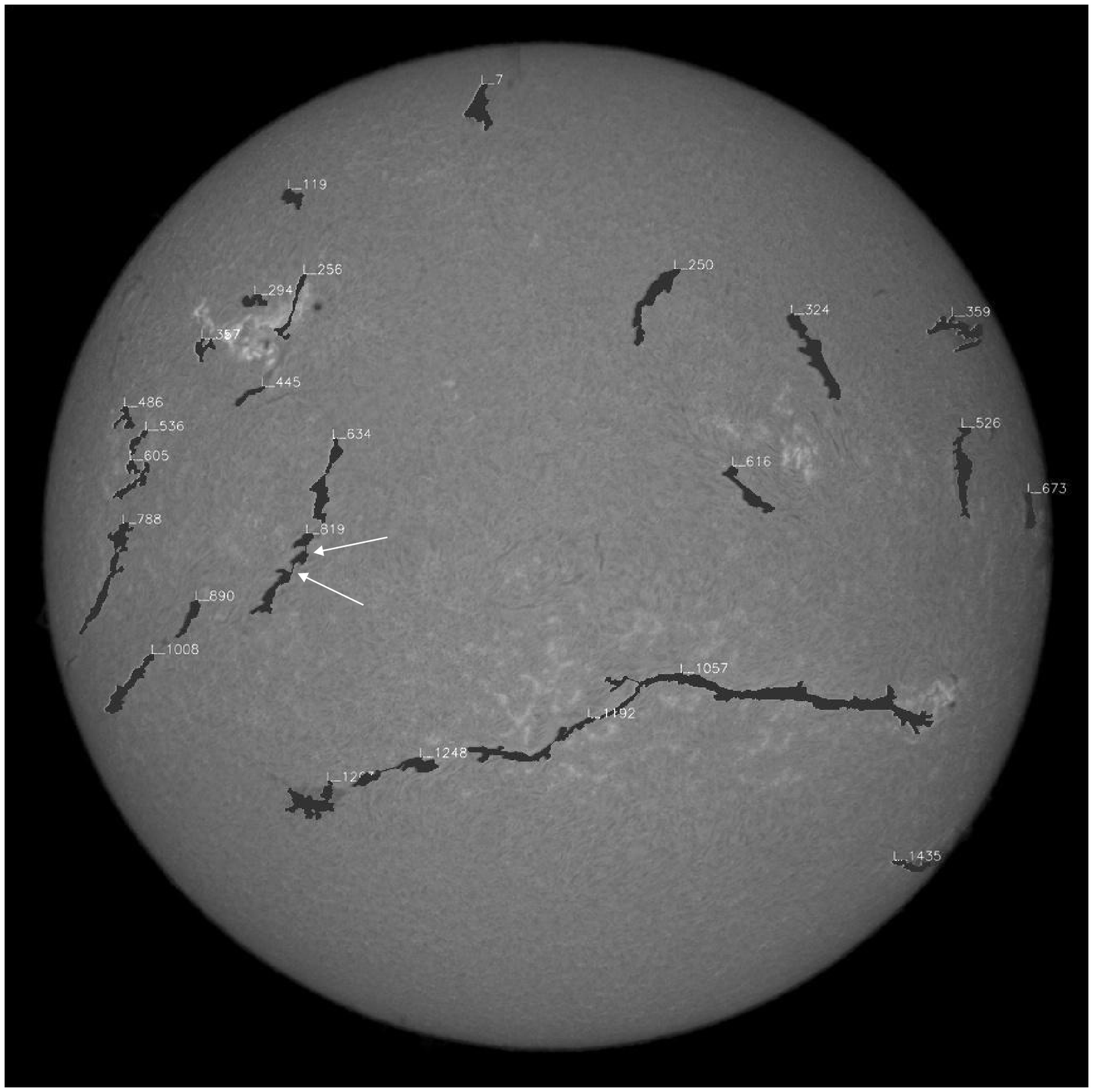}{fil_fig}
   {Sample KSO H$\alpha$ image from February 11, 2015 together with the filament 
   detections. Each filament is assigned and tracked by its unique ID (annotated 
   close to each filament). Individual filaments may can consist of more
   than one segment. Depending on their distance and direction they are 
   merged to a single ID, as an example see L\_819 (indicated by arrows).}

\subsubsection{Filament detection}

The image recognition algorithm can distinguish between flares, filaments, sunspots
and background ({\it i.e.}\  quiet-Sun regions). Information logs about filament 
position, length and ID are the second output of this algorithm. The detection of 
filaments is more sophisticated than that of flares as the brightness of filaments 
is not always clearly distinguishable from the background. In addition, parts of 
filaments may seem to vanish for short times due to degrading seeing conditions 
or due to filament oscillations \citep[{\it e.g.},][]{Hyder1966,Bi2014}. Filaments may also consist 
of more than one segment. In our algorithm, segments of a filament are connected 
together if they are oriented in the same direction and if they are separated by no 
more than $50''$. Figure~\ref{fil_fig} shows a sample H$\alpha$ image together with 
the filament detections. Each filament is identified by a unique ID that is 
tracked from each image to the next. In a future activity, the detected filaments 
and tracking of their ID in time will be used in order to implement an algorithm 
that automatically detects the eruption of filaments in the real-time data.

\subsection{Evaluation of the KSO real-time flare detection}

The KSO real-time flare and filament detection system went online in March 2013. 
After a testing and iteration period of about 4~month, it was fully established 
and runs continuously since about July 2013 in the same configuration. 
A detailed evaluation of the period between end of June 2013 and end of November 
2013 can be found in \cite{Poetzi2015}. In the following, we present an assessment 
of the real-time flare detection for an almost 2.5~year period from July 1, 2013 
until November 30, 2015. In total 824 flare have been detected during this period 
by the KSO automatic system. The flares detected are distributed among the 
importance classes as follows: 650 subflares (78.89\,\%), 
155 flares of class 1 (18.81\,\%), 17 flares of class 2 (2.06\,\%) and two flares 
of class 3 (0.24\,\%).\footnote{Note that we included only subflares with an area larger than 50~micro-hemispheres. Thus, in our distribution 
among the flare classes we have a larger fraction of flares of importance $\ge$1 than on average, cf.\ \cite{temmer2001}.}  
In order to evaluate the performance of system, we compare 
the real-time results obtained by the KSO automatic image recognition system 
against the official optical flare reports provided by NCEI (NOAA) and by KSO. NCEI (NOAA) 
and KSO flare reports both are obtained after the event occurrence by visual inspection of 
the data by experienced observers. 

Figure~\ref{diff_fig} shows the offsets for flare peak times, flare start times and
flare positions in heliographic latitude and longitude coordinates between the 
KSO automatic detections and the visual inspections. 89\,\% (95\,\%) of the flare 
peak times determined lie within $\pm$3~min ($\pm$5~min) of the values given in the official 
flare reports. For the start times, it is 65\,\% (76\,\%).
For the flare start time, the dispersion is somewhat larger and the distribution 
is asymmetric around zero with a tendency to positive values. This delay of the
automatic detections with respect to the flare reports probably result form the 
requirement that in the automatic algorithm at least three consecutive detections 
are necessary to define a brightening as a flare, whereas for visual inspection 
single images may be sufficient to detect the start of a flare. Large deviations 
(in both directions) are mostly a result of data gaps in the real-time image sequences.
The distributions of the heliographic flare positions, determined at the peak 
of the flare, are plotted in the bottom row of Figure~\ref{diff_fig}. They reveal 
a strong peak at values smaller than $1^\circ$. Basically all events detected lie 
within an accuracy of $\pm 5^{\circ}$ in heliographic latitude and longitude compared to the official flare reports based on visual inspection.

\articlefigurefour{\figspath/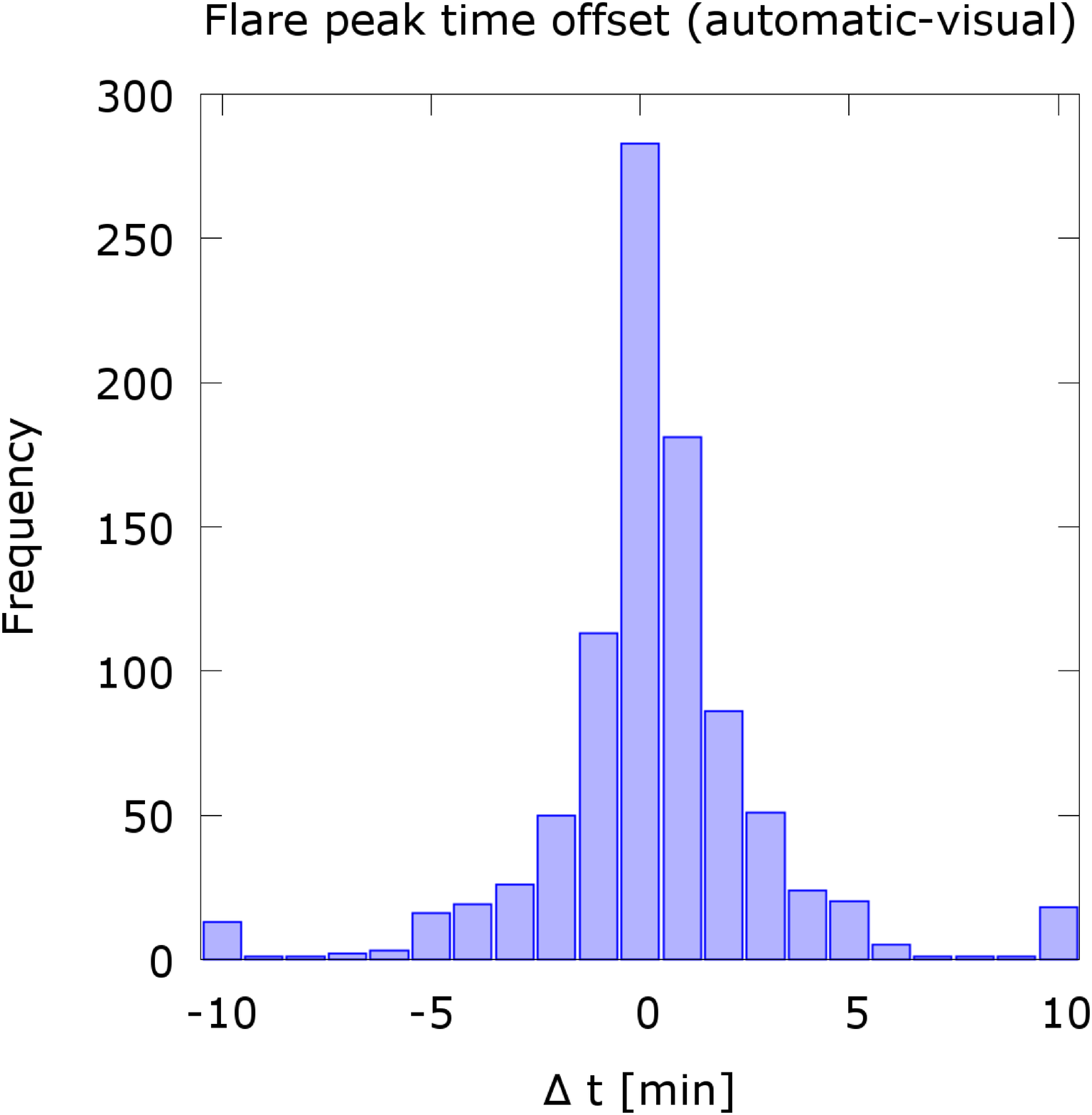}{\figspath/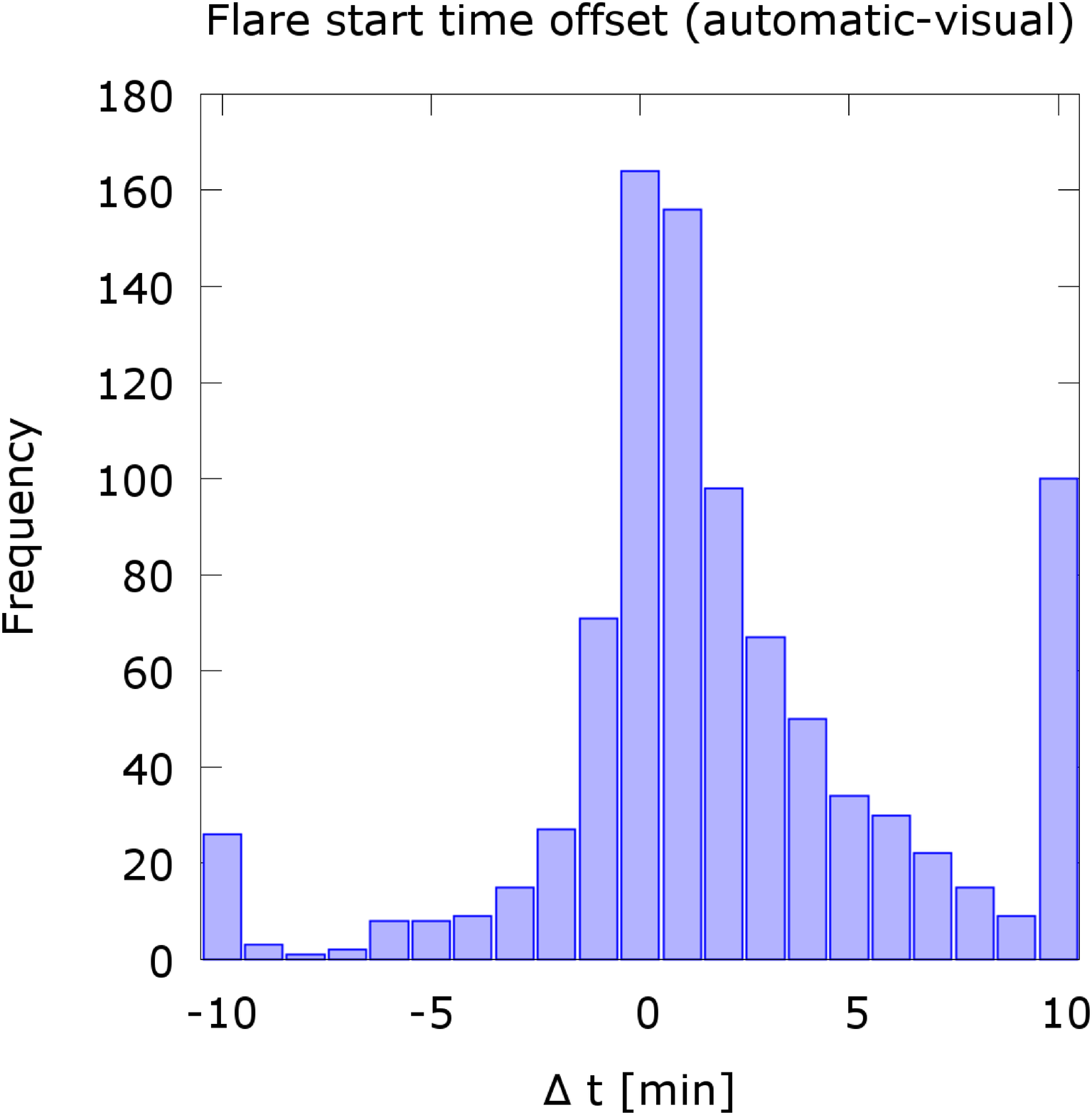}
   {\figspath/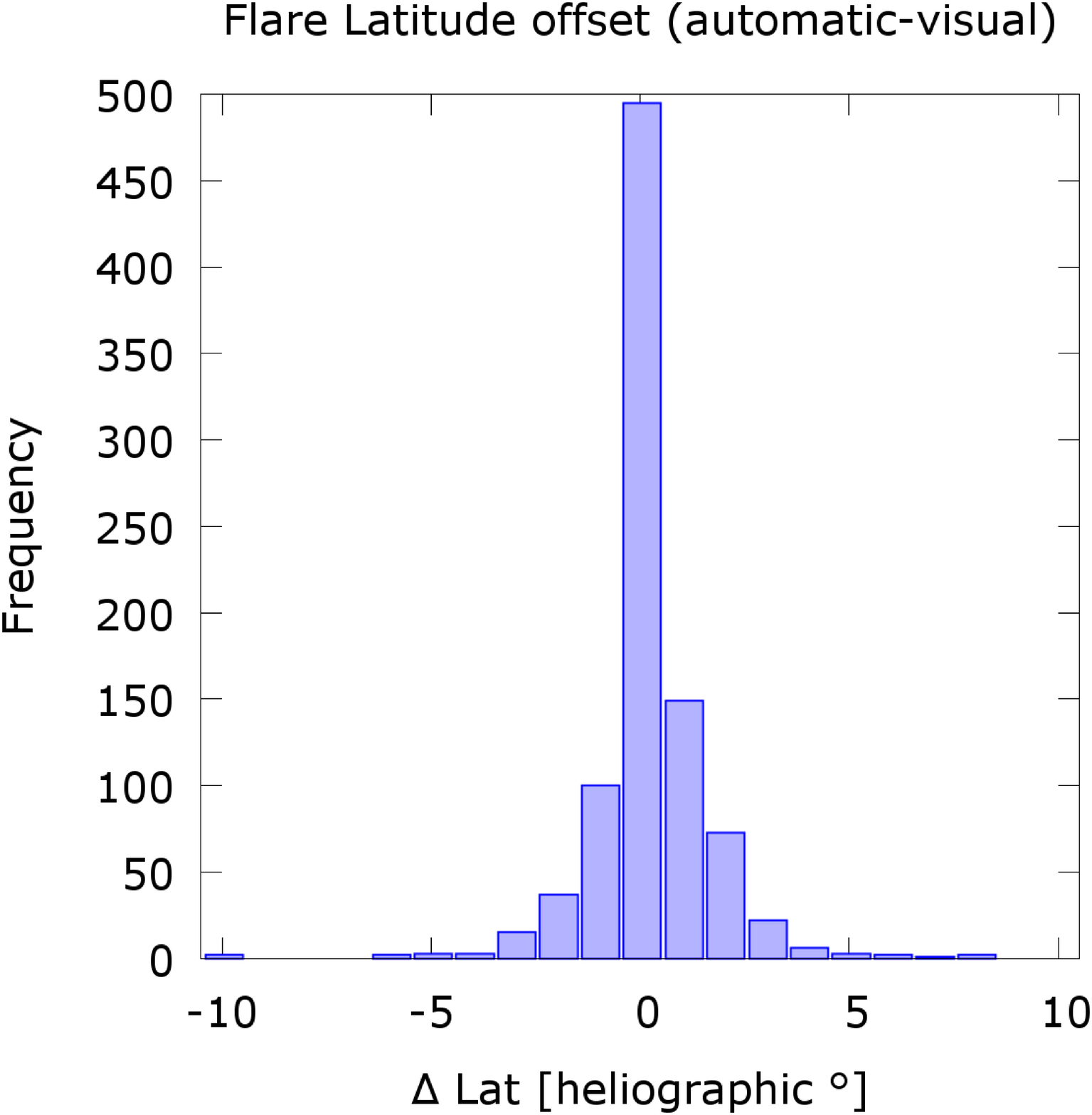}{\figspath/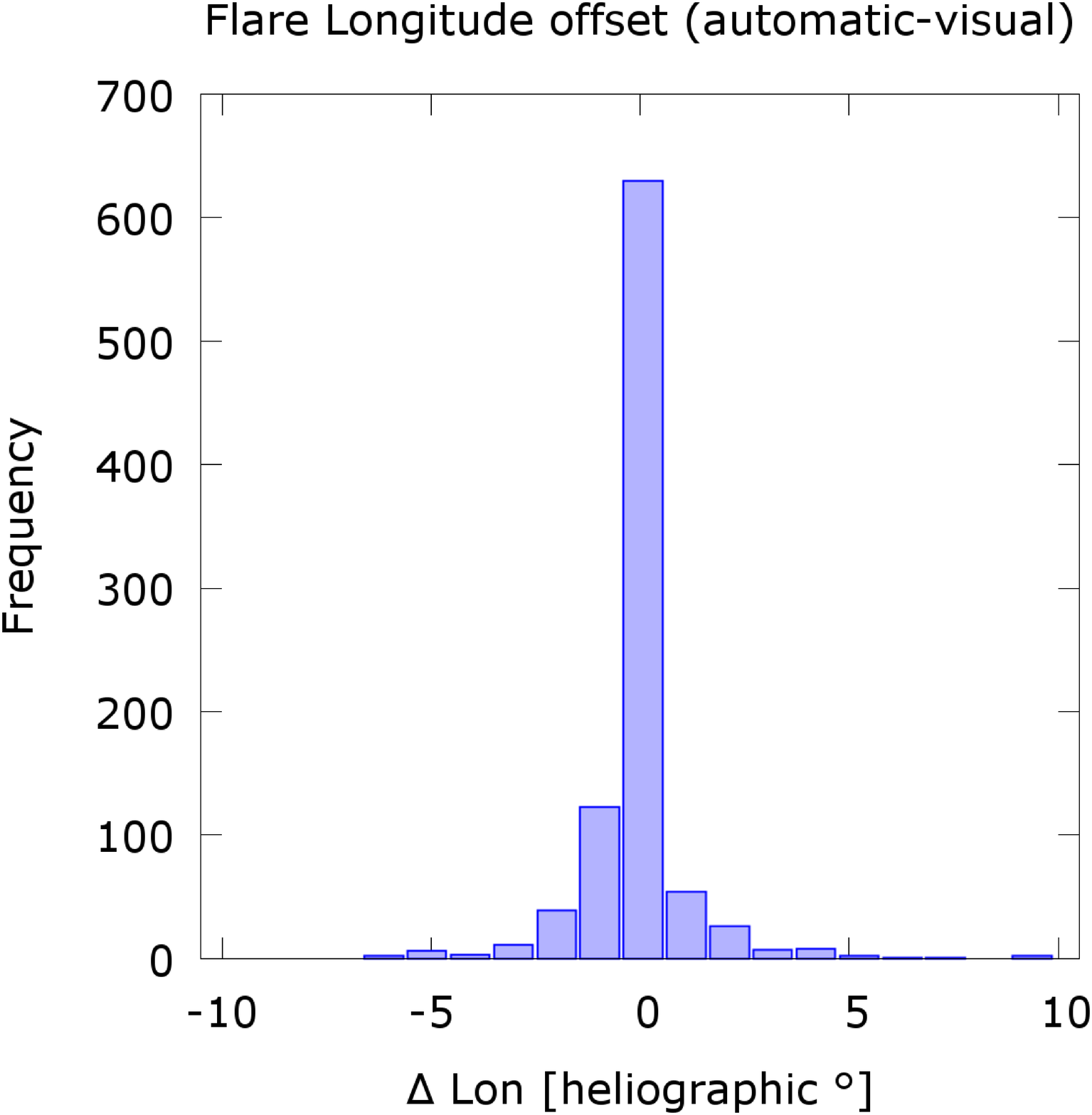}{diff_fig}
   {Offsets in detected flare peak times (\emph{top left}), flare start 
   times (\emph{top right}), flare latitudes (\emph{bottom left}) 
   and flare longitudes (\emph{bottom right}) between the KSO automatic real-time flare 
   detection system and the visual flare reports.
   }

In order to quantify the detection rate, we used the concept of verification measures
\cite[{\it e.g.},][]{Devos2014}. For this purpose we need four measures,
which we defined either (i) by comparing for each minute of observations the state of the automatic flare detection system 
with the official flare reports based on visual inspection ({\it continuous/ time series approach}) or (ii) by comparing the automatically detected flares 
with the flare reports ({\it discrete/ event approach}):
\begin{itemize}
  \item \emph{h} number of hits: automatic = flare AND visual = flare,
  \item \emph{f} number of false detections: automatic = flare AND visual $\ne$ flare,
  \item \emph{m} number of missed detections: automatic $\ne$ flare AND visual = flare,
  \item \emph{z} number of zeros: automatic $\ne$ flare AND visual $\ne$ flare.
\end{itemize}
This verification was applied to all flares that where larger than subflares, {\it i.e.}\ including all events of importance class 1, 2 and 3 
(no importance class 4 flare occurred in the evaluation period). For the continuous approach, the verification metrics obtained from this set are:  
\begin{eqnarray*}
  {\rm HR}  =  \frac{h+z}{h+f+m+z}  & = & 95\% \, ,\\
  {\rm POD}  =  \frac{h}{h+m}  & = &  87\% \, , \\
  {\rm TSS}  =  \frac{hz-fm}{(h+m)(f+z)}  & = &  0.82 \, .
\end{eqnarray*}
The hitrate HR gives a measure of the correct hits. However, we note that it is 
strongly influenced by the number of zeros ($z$), which is very high compared 
to the hits ($h$) --- this is due to the fact that flares are rare events. 
{\it E.g.}, in the year 2014, 95\% of the observation time were counted as zero 
({\it i.e.}\ no flare of importance $\ge$1 was ongoing). Therefore, the probability 
of detection (POD) is a more appropriate measure, as it does not include the zeros. 
It is still indicative of a high value of detection. The true skill score (TSS) 
or also called \emph{Hanssen and Kuipers Score} \citep{Hanssen1965}
gives information on the overall quality of the detection, considering both the true positive rates and false positive rates. 
The values of the TSS lie in the range $[-1,+1]$, where $+1$ means a perfect detection and $-1$ a perfect 
inverse detection. HSS values around 0 imply that the detection is no better than a random guess.
The obtained HSS value of 0.82 is high, indicating a good performance 
of the system. However, we note that the HSS is also biased by the large number
of zeros in the measurement series. Thus, evaluation quantities that are based on the 
events detected instead of the continuous measurement time series may provide a 
better quantification of the performance of the system.

If count the number of flares with an importance level of at least class 1 ({\it i.e.} ignoring subflares)
that were detected by the automatic system during the evaluation period,  we arrive at a total of 174 events . However, 
28 out of these were false alerts. (21 of these false alerts can be attributed 
to data gaps, which led to a splitting-up of ongoing flares in the automatic routine.)
From these numbers we obtain for the discrete detection of flare events of class 1 and 
larger a false alarm rate of
\begin{eqnarray*}
  {\rm FAR}  =  \frac{f}{h+f}  & = &  16.1\% \,.
\end{eqnarray*}
In addition, 8 events that were classified as importance~1 or larger by the NOAA or 
KSO optical flare reports, were not correctly identified by the automatic system. 
They were detected as flares but assigned as subflares. Thus, for the discrete verification approach, we obtain for the detection 
of flares of importance class $\ge$1 a probability of detection (POD) of 94.8\%.

\section{Conclusions and Outlook}

Ground-based observations provide an important means to monitor and to study the solar sources of
space weather disturbances, complementing the data obtained from space-based observatories. 
The main advantages of ground-based observations are the comparatively simple and non-expensive observing systems, the flexibility in upgrading the system, fast networks and data transfer,
and the immediate availability of the data. These facts are highly relevant for the real-time assessment of space weather events. 
Difficulties that arise from ground-based observations are data gaps and image degradation due to the day/night-cycle, local weather and seeing conditions. 
However, these drawbacks can in principle be overcome by networks of observatories densely distributed over the world. Currently, there exist two networks that provide high-cadence full-disk H$\alpha$ observations of the Sun 
for space weather monitoring and research, GONG (consisting of six sites well distributed in longitude, with same instrumentation) and GHN (nine sites, different instruments). 

In this review, we also presented an assessment of the real-time detection system of solar flares and filaments in H$\alpha$ images that was established in 2013 at Kanzelh\"ohe Observatory,
and that is continuously running since then. For the evaluation period July 2013 to November 2015, the discrete detection probability for flares of H$\alpha$ importance $\ge$1 was 95\,\%, with a 
false flare alarm rate of 16\,\%. We note that the other 5\% of events were also detected but erroneously classified as subflares. The false alarms were mostly resulting from data gaps, which led to an assignment of individual flares as two or multiple events by the automatic detection system. Currently, there is an effort to utilize the automatically identified filaments in order to develop a system for the real-time detection of filament eruptions, which are often accompanied by CMEs. Further improvements could be achieved, when the real-time flare and filament detection system was extended to a network of observatories, thus filling up data gaps from one station by data from another station. Major challenges will be to test whether robust image recognition algorithms can be developed that work successfully on sequences of data that are obtained 
by different instruments (in terms of telescope, filter, CCD camera etc.) or whether homogeneous data sets like from the GONG network are needed. 


\acknowledgements This study was developed within the framework of ESA Space Situational 
Awareness (SSA) Programme (SWE SN IV-2 activity).

\bibliography{asv_literature}  

\end{document}